\documentstyle[11pt,twoside,epsf,newpasp]{article}
\begin{document}
%Version 2.0.0
\title{On the association of G343.1-2.3 and PSR 1706-44}
\author{Dodson, R.}
\affil{University of Tasmania, Australia}
\author{Golap, K.}
\affil{NRAO, USA}
\author{Osborne, J.}
\affil{University of Durham, UK}
\author{UdayaShankar, N.}
\affil{Raman Research Inst., India}
\author{Rao, A.P.}
\affil{National Centre of Radio Atron.,India}
%\maketitle
\begin{abstract}
The observations of the supernova remnant G343.1-2.3 with the Mauritius
Radio Telescope, the Australia Compact Array, and the Hobart Single
dish are presented. With these more sensitive measurements the
association with the pulsar PSR1706-44 becomes much more likely.  The
major points from the observations are presented.
\end{abstract}

\section{Introduction}
The pulsar PSR B1706-44 is one of only 7 that are known to emit
gamma-rays in the GeV energy range and one of only 3 that have been
detected in the TeV range.  There is thus considerable interest in the
circumstances of the origin and evolution of such high energy pulsars.

The radio pulsar was discovered by Johnston et al (92). McAdam et al
(93) published a map made by the MOST telescope at 843MHz of the area
around the source. It showed a semicircular arc of emission, which has
subsequently been denoted as the supernova remnant (SNR) G343.1-2.3,
with the pulsar seemingly embedded in a small feature at its south
eastern extremity.  The $\Sigma-D$ distance is 3~kpc for the remnant.
The pulsar distance indicated by the Taylor and Coordes model was only
1.8~kpc. Frail et al (94) and Nicastro et. al. (96) both argued
against the associations.

\section{The Importance of Associations}

Incorrect, or chance, associations can bias
statistical results, especially when the number of objects is
small. Pulsar and supernova remnant associations have been examined
critically by Kaspi (96).

This pulsar, has a characteristic age of about 17.5~kyears (Johnston
et al 92), a HI absorption distance of 2.4-3.2~kpc (Korbalski 95) and
a scintillation velocity of $<=$ 100~kms$^{-1}$ (Dodson 97). The SNR
G323.1-2.3 has an age of between 8 and 17~kyears, and a distance of
3-3.6~kpc (Dodson 97). The pulsar sits about $2/3^{rds}$ from the
geometric centre requiring a velocity of 750~kms$^{-1}$.

The Taylor and Coordes model, 1.8~kpc, is clearly in error. The values
for the scintillation bandwidth to velocity have been recalculated to
reflect the improved models (Dodson 97). The shortfall in the
scintillation velocity is not of great concern as it is within the 2
$\sigma$ limit of the distribution of measured velocity to
scintillation velocity, furthermore the geometric centre is rarely the
true SN site.

Altering Kaspi's scoring method (96) slightly, (justified in detail in
Dodson 97), these change the previous score of -3 to one of +2. These
are summarised in table \ref{tab:scores}.

\begin{table}
\begin{center}
\begin{tabular}{|l|l|l|l|l|}
\hline
Age (kyr) & Distance (kpc) & $\beta$ & $v_t$ (km/s) & Score\\
\hline
\multicolumn{5}{|c|}{Previous PSR to SNR values}\\
\hline
17.5/? &  1.8 / 3 & 1.0 & $<$50 / 1000 & -3 \\
\hline
\multicolumn{5}{|c|}{Now}\\
\hline
17.5/8-17 & 2.4-3.2 / 3-3.6 & 0.7 & $<$100 / 750 & 2 \\
\hline
\end{tabular}
\caption{Pulsar and SNR parameters}
\label{tab:scores}
\end{center}
\end{table}

\section{Observations of G323.1-2.3}
 
The Mauritius Radio Telescope (MRT) (Golap et al 98) has observed this
supernova remnant at 151.5~MHz, and a resolution of $4' \times
4'.2$. I present preliminary results, with a resolution of $4' \times
15'$. The MRT is particularly suited for non-thermal (as it is low
frequency) and broadscale sources (as it is a filled {\em uv}
instrument).

Follow-up observations were done with the Australian Telescope Compact
Array (ATCA), and the Mount Pleasant 26m single dish instrument. These
were combined to produce the the final maps.

\begin{figure}
\begin{center}
\plottwo{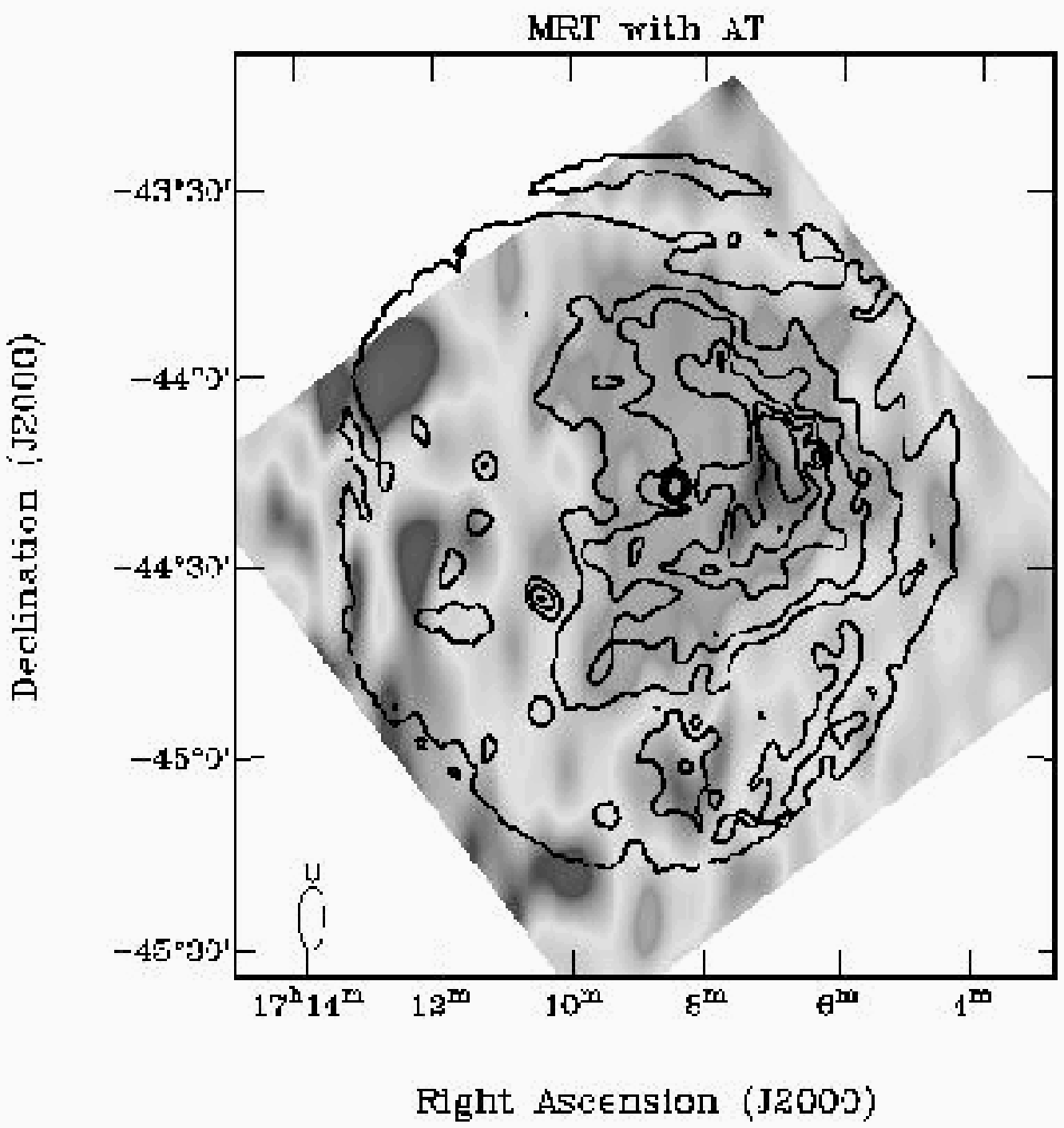}{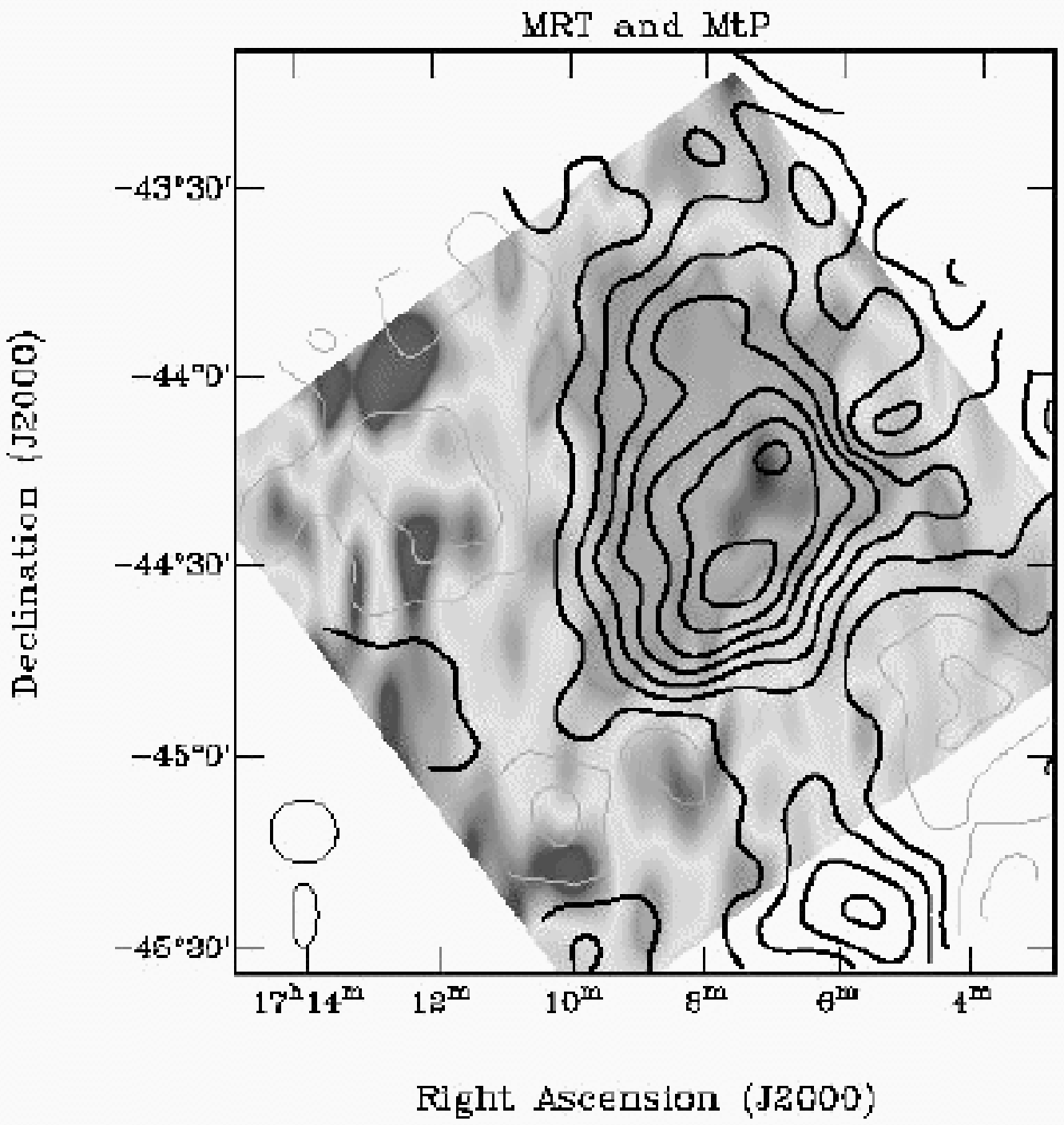}
\caption{MRT overlayed with a) ATCA and b) Mt Pleasant}
\label{fig:MRT+AT}
\end{center}
\end{figure}

These show that the observations with MOST, the ATCA and VLA are under
estimating both the received flux and source extent.

\section{The spectral index}

The spectral index of this source is the clearest evidence of the
problem with the other observations. The {\em single dish}
observations show a monotonic spectral index of 0.5 However the MOST,
VLA and ATCA--only observations find a significant short fall in flux
and (more importantly) extent. It is only when the single dish
information is included in the interferometer images that sensible
results can be obtained.

\section{Techniques for image recovery}

There is no replacement for making the observation.  The total power
is the most important measurement if one is not to under estimate the
flux from and the extent of the source.  This is best done with a
single dish observation.  For the ATCA, the dishes are 22m in diameter
and and a closest approach of 30m. Therefore the Parkes 64m dish is an
ideal {\em uv} filler.  However it is also 10 times over subscribed,
so Mount Pleasant, with a diameter of 26m, is quite adequate.

It is gratifying to note that MRT, an interferometer, also includes the
shortest possible spacing and thus recovers the broadscale structure
naturally.

\section{The ROSAT X-ray observations}

\begin{figure}
\begin{center}
\plotone{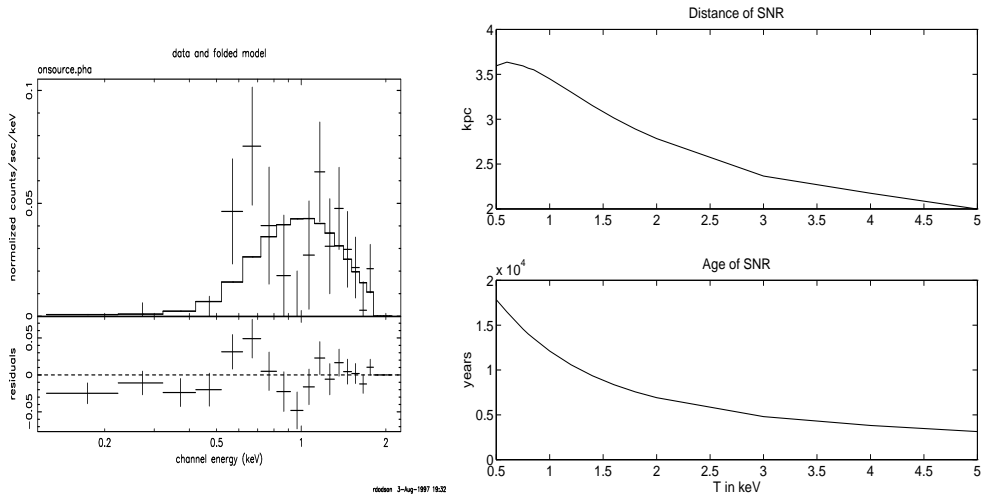}
\caption{ROSAT residual counts, and fitted distance \& age model}
\label{fig:rosat}
\end{center}
\end{figure}

As one can see, the data is being pushed hard; the background
subtraction is distorted by the LMXB 1705-44 and the poor PSF of
ROSAT. Full details of how the background subtraction and model fitting
were done can be found in Dodson (97). Fortuitously, the distance, but
not the age, (Figure \ref{fig:rosat}) is robust for the range of
possible model fits to be 3--3.6~kpc. The age can only be estimated as
between 8--17~kyears.

\section{conclusions}

The association has been shown to be highly likely, with good alignment
in both time and space. The uncertainties in the X-ray model will be
tackled with the new generation of X-ray observatories. The question of
the pulsar proper motion will be measured as soon as we have a phase
reference.

\end{document}